\documentclass[journal,twoside]{IEEEtran}
\usepackage{cite}
\usepackage{multirow}
\usepackage{graphicx}
\usepackage{placeins}
\usepackage[caption=false, font=footnotesize]{subfig}
\usepackage{amsmath}
\usepackage{color}
\usepackage[usenames, dvipsnames]{xcolor}
\usepackage{soul}
\usepackage{array}
\usepackage{url}
\usepackage{comment}

\begin{document}

\title{Time-Interleaved C-band Co-Propagation of Quantum and Classical Channels}
\author{
    Jing Wang,~\IEEEmembership{Member,~IEEE,}
    Brian J. Rollick,~\IEEEmembership{Member,~IEEE,}\\
    Bernardo A. Huberman,~\IEEEmembership{Fellow,~American Physical Society}
    \thanks{Manuscript received September XX, 2023; revised XXX XX, 2023.}
    \thanks{J. Wang, B. J. Rollick, and B. A. Huberman are with CableLabs, Louisville, CO, 80027. Email: j.wang@cablelabs.com}
    \thanks{J. Wang and B. J. Rollick contributed equally to this work.}
    }

\markboth
    {Journal of Lightwave Technology,~Vol.~xx, No.~xx, XXX~2023}
    {Wang \MakeLowercase{\textit{et al.}}: Time-Interleaved C-band Co-propagation of Quantum and Classical Channels}
\maketitle

\begin{abstract}
A successful commercial deployment of quantum key distribution (QKD) technologies requires integrating QKD links into existing fibers and sharing the same fiber networks with classical data traffic. To mitigate the spontaneous Raman scattering (SpRS) noise from classical data channels, several quantum/classical coexistence strategies have been developed. O-band solutions place the QKD channel in the O-band for lower SpRS noise but with the penalty of higher fiber loss and can rarely reach beyond 80 km of fiber; another method is C-band coexistence with attenuated classical channels, which sacrifices the performance of classical channels for lower SpRS noise. In this work, a time-interleaving technique is demonstrated to enable the co-propagation of quantum and classical channels in the C-band without sacrificing either performance. By embedding QKD pulses in the gaps between classical data frames, the quantum channel is isolated from SpRS noise in both wavelength and time domains. C-band co-propagation of a polarization-encoding decoy-state BB84 QKD channel with a 100 Gb/s QPSK channel is experimentally demonstrated with quantum bit error rate (QBER) of 1.12\%, 2.04\%, and 3.81\% and secure key rates (SKR) of 39.5 kb/s, 6.35 kb/s, and 128 b/s over 20, 50, and 100 km fibers, respectively. These results were achieved with the presence of classical launch power up to 10 dBm, which is at least one order of magnitude higher than reported works. We also demonstrated the co-propagation of a QKD channel with eight classical channels with total launch power up to 18-dBm (9-dBm per channel), which is the highest power of classical channels reported in C-band coexistence works.

\end{abstract}


\section{Introduction}
\IEEEPARstart{S}{ECURITY} of today’s cryptographic algorithms is based on computational complexity, which is being challenged by the emergence of quantum computers running Shor’s and Grover’s algorithms~\cite{Grover_PRL_97, Shor_SIAM_99, Google_Nature_19}. To deal with the threats from quantum computers, post-quantum cryptographic (PQC) algorithms have been developed. The National Institute of Standards and Technology (NIST) has organized several rounds of competitions for PQC algorithms, but their security is still questionable. In 2022, two finalists of NIST competitions, post-quantum signature scheme Rainbow and Supersingular Isogeny Key Encapsulation (SIKE) were broken~\cite{Beullens_22, Castryck_22}.

\begin{figure*}[ht]
    \centering
    \includegraphics[width=1\linewidth]{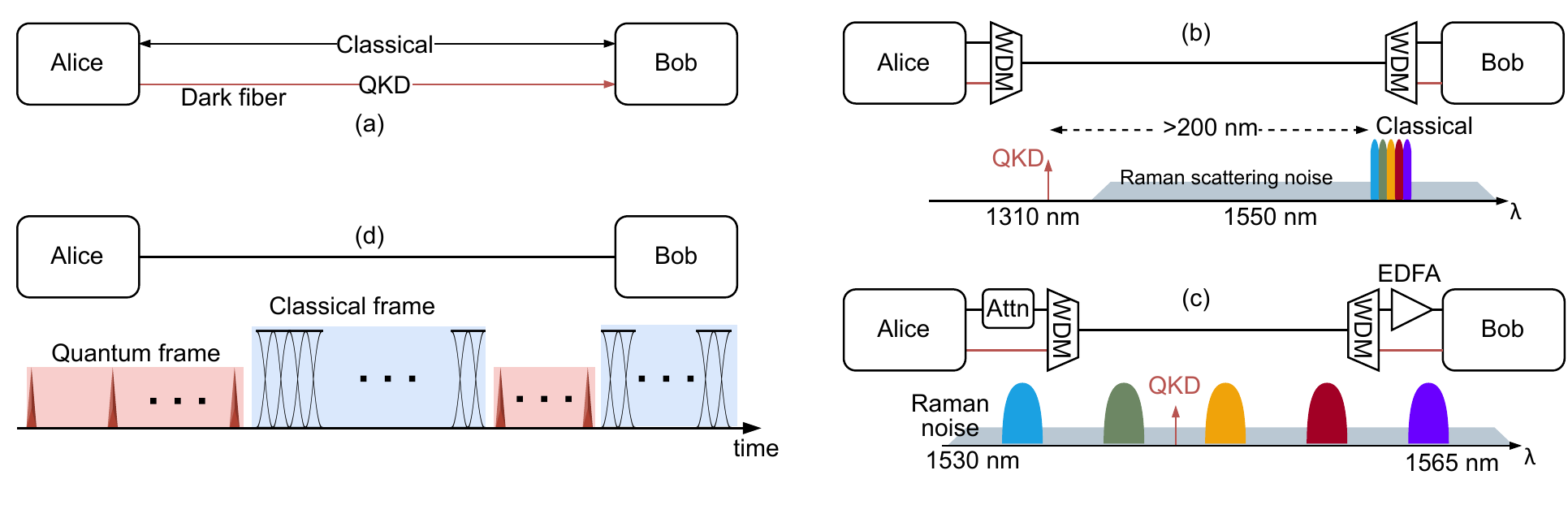}
    \caption{Existing solutions to the coexistence of quantum and classical channels in a shared fiber. (a) Dedicated dark fiber. (b) Quantum channel in the O-band. (c) Attenuate classical channels. (d) Time-division multiplexing of quantum and classical data frames.}
    \label{fig:exist}
\end{figure*}

Quantum key distribution (QKD), on the other hand, provides information-theoretic security~\cite{Bennett_Science_92, Gisin_RevModPhys_02, Scarani_RevModPhys_09, FH_Xu_RevModPhys_20} since the confidentiality of key bits is guaranteed by quantum mechanics. So far, most QKD research focuses on extending distances~\cite{} or increasing key rates~\cite{Toshiba_Nature_18, Toshiba_NaturePhotonics_21, Yuan_NatureComm_23, Pan_NaturePhotonics_23}. Only a few innovations have been done from the deployment perspective. Most reported QKD systems require dedicated/dark fibers similar to Fig.~\ref{fig:exist}(a) since the spontaneous Raman scattering (SpRS) noise from classical data channels would easily overwhelm a QKD link. However, in today's telecommunication networks, fibers are scarce and expensive resources. It is cost-prohibitive for network operators to add new fibers or reserve them for only quantum purposes, so integrating QKD into existing fibers and sharing the same fiber networks with classical data traffic is an essential requirement for the commercial success of QKD technologies.

\begin{figure*}[ht]
    \centering
    \includegraphics[width=0.75\linewidth]{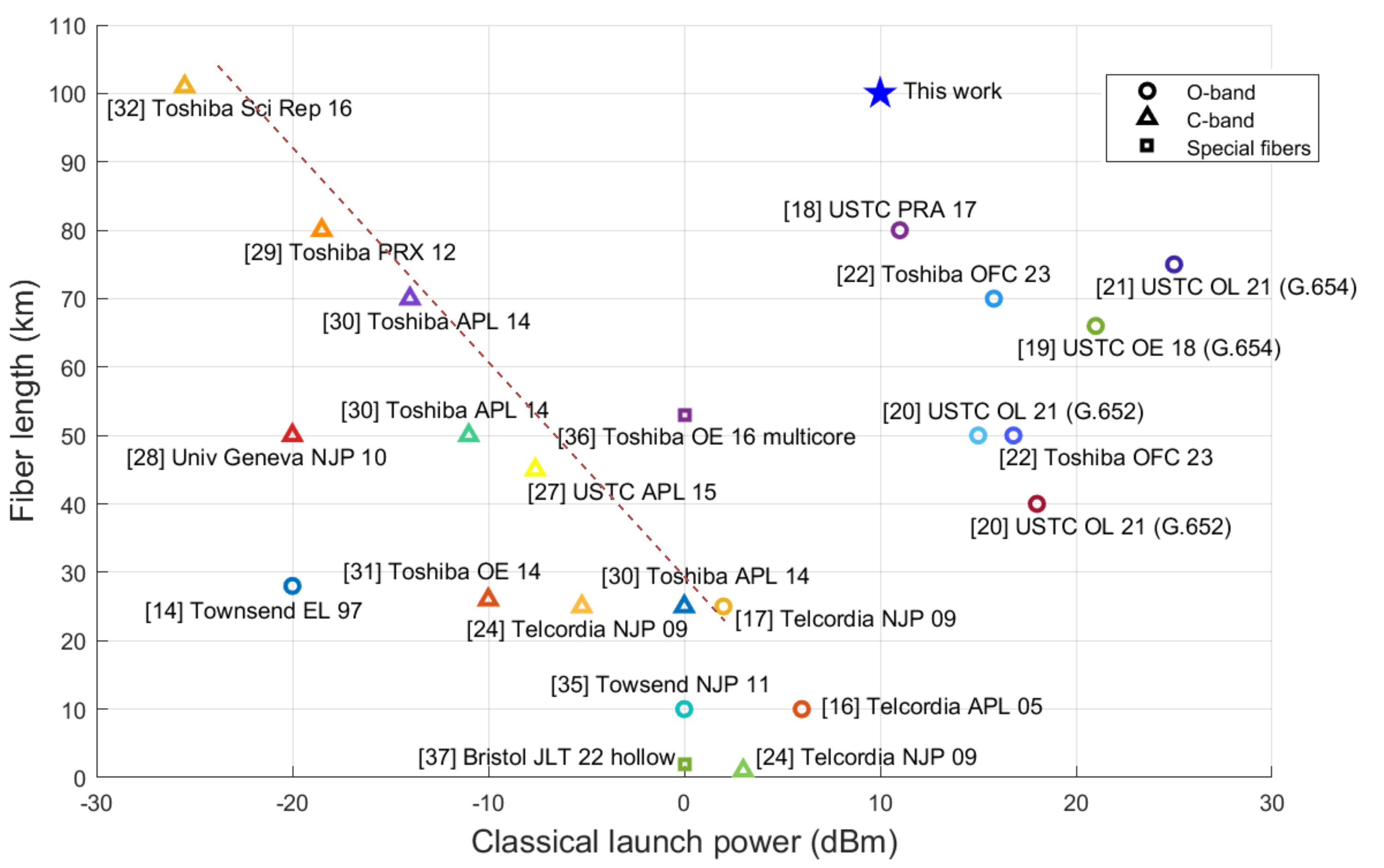}
    \caption{State-of-the-art in terms of fiber distances and classical launch power for QKD/classical coexistence in the same fiber.}
    \label{fig:state_art}
\end{figure*}

Several quantum/classical coexistence technologies have been developed with a focus on eliminating the SpRS noise from classical channels. Townsend first proposed the wavelength division multiplexing (WDM) of quantum and classical channels by placing the QKD channel in the O-band and classical data traffic in the C-band~\cite{Townsend_EL_97}, shown in Fig.~\ref{fig:exist}(b). The feasibility of this method was first proven by using continuous-wave (CW) lasers to emulate classical data traffic~\cite{Telcordia_LEOS_03, Telcordia_APL_05, Telcordia_NJP_09_1310}. Thanks to the large wavelength separation, the SpRS noise in the O-band is orders of magnitude smaller than that in the C-band~\cite{USTC_PRA_17}. Coexistence with terabits classical data traffic with more than 20-dBm launch power has been demonstrated~\cite{USTC_OE_18, USTC_Guo_OL_21_SMF, USTC_Guo_OL_21_ULL, Toshiba_OFC_23}. Fig.~\ref{fig:state_art} shows the state-of-the-art of quantum/classical coexistence in terms of fiber distances of QKD channels and the maximum allowed launch power of classical channels. O-band results are labeled by circles. Most world-record results were achieved by placing QKD channels in the O-band, e.g., the highest classical data rate of 7.168 Tb/s~\cite{USTC_PRA_17}, longest fiber distance of 80 km~\cite{USTC_PRA_17}, largest classical launch power of 25 dBm~\cite{USTC_Guo_OL_21_ULL}, and maximum co-propagation efficiency of $\mathrm{253.7 Mb/s\cdot mW \cdot km}$~\cite{Toshiba_OFC_23}. However, this method is limited by the high fiber loss in the O-band and can rarely reach beyond 80 km~\cite{USTC_PRA_17}, even using G.654 ultra-low loss fibers~\cite{USTC_OE_18, USTC_Guo_OL_21_ULL}. Network compatibility is another issue since most deployed optical routers only support C-band operations.

\begin{figure*}[ht]
    \centering
    \includegraphics[width=0.8\linewidth]{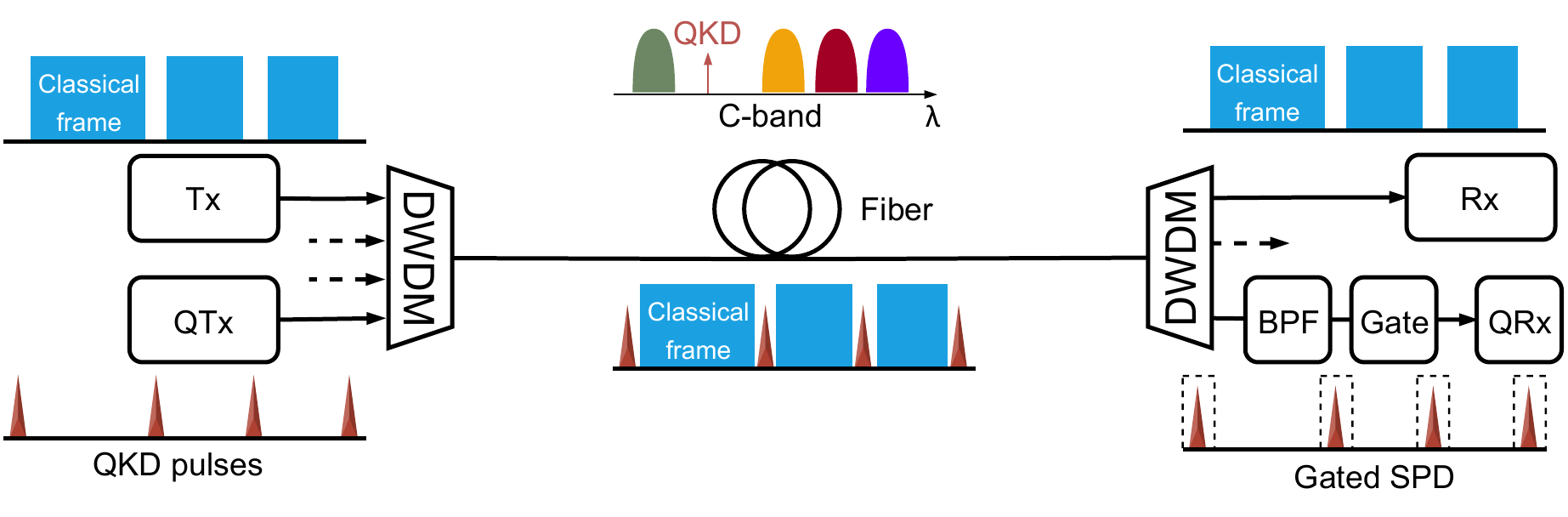}
    \caption{Time-interleaving of QKD pulses and classical data frames. QTx/QRx: Quantum transmitter/receiver. BPF: bandpass filter.}
    \label{fig:operation}
\end{figure*}

Another method is C-band coexistence with attenuated classical channels~\cite{Qi_NJP_10}, which sacrifices the performance of classical channels for lower SpRS noise, shown in Fig.~\ref{fig:exist}(c). CW lasers were used in early works to emulate classical data channels~\cite{Telcordia_NJP_09_1550, Telcordia_OFC_10, USTC_OE_10}, then scaled to 100 Mb/s \cite{USTC_APL_15},  ~1 Gb/s \cite{Univ_Geneva_NJP_10, Toshiba_PRX_12} and 10 Gb/s \cite{Toshiba_APL_14, Toshiba_OE_14}. In these works, the classical channels were attenuated to match the receiver sensitivities. This method only works for low data rate communications since there is plenty of power margin for classical channels. For 100 Gb/s and beyond, normal QKD operation is impossible unless classical channels are attenuated below receiver sensitivities~\cite{Toshiba_SciRep_16}. In some works, classical channels were first attenuated heavily and then amplified after fiber propagation before the receiver~\cite{Toshiba_OE_14, Toshiba_SciRep_16}. This method would fail to meet the distance and bit error rate (BER) requirements in real networks. In Fig.~\ref{fig:state_art}, all C-band results are labeled by triangles and below a dashed line, which shows a trade-off between the fiber reach of QKD channels and the launch power of classical channels. Compared with O-band solutions, C-band results can reach longer fiber distances but have strict limits on launch power, channel number, and data rates of classical channels. 

A third method is time-division multiplexing (TDM) of quantum and classical data frames~\cite{Cisco_PhysRevRes_22, UCDavis_OFC_23}. It was initially designed to enable packet switching of quantum payloads by adding classical headers and trailers before and after the quantum payload, shown in Fig.~\ref{fig:exist}(d). This method allows quantum and classical packets to access the same fiber alternatively but in a low-efficiency way. This is because QKD pulses are quite sparse in the time domain with narrow pulse widths and long periods. For example, a pulse train with 100-ps pulse width and 25-MHz repetition rate has a duty cycle of only 0.25\%. The remaining 99.75\% time slots between two consecutive pulses are empty. To employ time resource efficiently, Townsend proposed to send QKD pulses during the time slots of 0 bits of a co-propagating classical data stream and demonstrated QKD over a 10-km fiber with -2.7 dBm launch power~\cite{Townsend_NJP_11}. Other coexistence solutions include the utilization of special fibers, such as multicore~\cite{Toshiba_OE_16} and hollow fibers~\cite{Bristol_JLT_22}, labeled by squares in Fig.~\ref{fig:state_art}, which are subjected to high manufacturing costs and short distances.

In this work, a time-interleaving technique is developed to enable the co-propagation of quantum and classical channels in the C-band of the same fiber with the highest reported launch power of classical channels. By placing QKD and classical channels on different wavelengths and embedding QKD pulses into the gaps between classical data frames, the quantum channel can be isolated from SpRS noise in both wavelength and time domains. This method leverages the low fiber loss and network compatibility of the C-band, while at the same time removing the power limit on classical channels, allowing co-propagation without sacrificing the performance of either classical or quantum channels. We experimentally demonstrate the C-band co-propagation of a polarization-encoding decoy-state BB84 QKD channel with a 100 Gb/s quadrature phase shift keying (QPSK) channel with quantum bit error rate (QBER) of 1.12\%, 2.04\%, and 3.81\% and secure key rates (SKR) of 39.5 kb/s, 6.35 kb/s, and 128 b/s over 20, 50, and 100 km fibers, respectively. These results were achieved with the presence of classical channel power of up to 10 dBm, which is one order of magnitude higher than other reported results. The relatively low SKRs are limited by the slow response and long dead time of our low-cost single-photon detectors (SPDs). Coexistence with eight classical channels is also achieved with a total launch power of up to 18 dBm (9 dBm per channel). To our knowledge, this is the highest power of classical channels in C-band coexistence works. In Fig.~\ref{fig:state_art}, all C-band results are below the dashed line due to the trade-off between the QKD distance and the power of classical channels. Our result (labeled by a blue star) is the only outlier above this distance-power limit.

This paper is organized as follows. Section~\ref{Op} shows the operation principles of the proposed time-interleaving technique. Section~\ref{exp} describes the experimental setup. Section~\ref{Results} presents the experimental results. Section~\ref{Dispersion} investigates the dispersion walk-off between QKD and classical channels. Finally, section~\ref{Conclusions} concludes the paper.

\section{Operation Principles}\label{Op}
Fig.~\ref{fig:operation} shows the operation principles of the time-interleaving technique. The QKD and classical channels use different wavelengths in the C-band. Since QKD pulses have a very low duty cycle, they could be embedded into the gaps between classical data frames. At Alice, the classical transmitter (Tx) and quantum transmitter (QTx) are synchronized so that QKD pulses and classical data frames are interleaved in the time domain. After fiber propagation, quantum pulses are separated from classical data frames in both wavelength and time domains. A bandpass filter (BPF) blocks the out-of-band SpRS noise from entering the quantum receiver (QRx) and the QRx is gated to block the out-of-window noise.

\begin{figure*}[ht]
    \centering
    \includegraphics[width=0.8\linewidth]{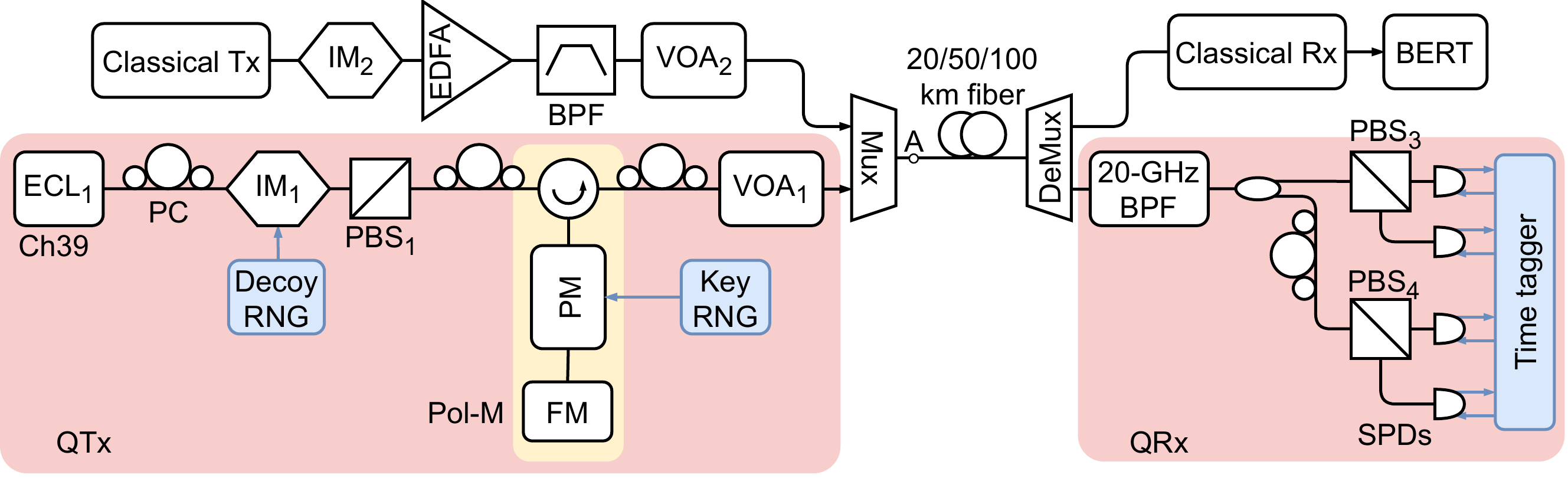}
    \caption{The experimental setup of coexistence of a polarization-encoding decoy-state BB84 QKD channel with classical channels. ECL: external cavity laser. PC: polarization controller. IM: intensity modulator. PBS: polarization beam splitter. PM: phase modulator. FM: Faraday mirror. Pol-M: polarization modulator. RNG: random number generator. VOA: variable optical attenuator. BPF: bandpass filter. SPD: single photon detector. EDFA: erbium-doped fiber amplifier. BERT: bit error rate test.}
    \label{fig:exp}
\end{figure*}

\section{Experimental Setup}\label{exp}

The experimental setup is shown in Fig.~\ref{fig:exp}. A polarization-encoding decoy-state BB84 QKD system with three intensities~\cite{BB_TheoCompSci_84} is shown in red boxes, whereas classical systems in blue. An external cavity laser (ECL) at 193.9 THz (ITU-T Ch39, 1546.12 nm) is used as the light source, followed by an intensity modulator ($\mathrm{IM_1}$) for pulse generation and decoy state preparation. A 10-GSa/s Tektronix arbitrary waveform generator (AWG) drives $\mathrm{IM_1}$ to generate 200-ps pulses at a repetition rate of 25 MHz. Thus, the QKD pulses have a 40-ns period and 0.5\% duty cycle. The polarization modulator (Pol-M) consisting of a circulator, a phase modulator (PM), and a Faraday mirror (FM), is used~\cite{Calgary_NJP_09, Toronto_PRL_14, Tennessee_PRA_21} to encode polarizations in two non-orthogonal conjugate bases, rectilinear and diagonal. The PM is driven by a Keysight function generator to prepare four polarization states $0^{\circ}$ (H), $45^{\circ}$ (D), $90^{\circ}$ (V), and $-45^{\circ}$ (AD) by applying voltages of 0, $\mathrm{V_\pi/2}$, $\mathrm{V_\pi}$, and $\mathrm{3V_\pi/2}$, where $V_\pi$is the half-wave voltage. The polarization beam splitter (PBS) before the Pol-M ensures a linear input polarization. A variable optical attenuator ($\mathrm{VOA_1}$) is adjusted to reduce the pulse intensity at point A to the single-photon level.

Two kinds of classical channels are tested in the experiments, a 10 Gb/s on-off keying (OOK) link, and a 100 Gb/s coherent QPSK link. The 100 Gb/s QPSK link is implemented by a pair of Acacia (Cisco) CFP2-DCO coherent transceivers. The classical channels have tunable wavelengths in the C-band to investigate the wavelength dependence of SpRS noise. To emulate the gaps between classical data frames, $\mathrm{IM_2}$ carves out windows on continuous data traffic to accommodate the QKD pulses. The gap window is 15 ns wide with a repetition rate of 25 MHz, synchronized with the 40-ns period of QKD pulses. Although 15 ns is more than enough to accommodate 200-ps QKD pulses, this wide window allows our further investigation of dispersion walk-off between quantum and classical channels. Due to dispersion, wider windows will be required for longer fiber distances and larger wavelength differences between classical and quantum channels. Besides a single classical channel, we were able to time-interleave the QKD channel with up to 8 classical channels without significant SpRS noise.

An erbium-doped fiber amplifier (EDFA) and $\mathrm{VOA_2}$ control the launch power of classical channels. A BPF after the EDFA eliminates the broadband amplified spontaneous emission (ASE) noise. In experiments, classical launch power at point A can be up to 10 dBm for a single channel or 18 dBm for eight channels. This classical launch power is at least one order of magnitude higher than reported C-band coexistence results~\cite{Univ_Geneva_NJP_10, Toshiba_PRX_12, Toshiba_APL_14, Toshiba_OE_14, Toshiba_SciRep_16}. Three fiber distances are tested in experiments, 20, 50, and 100 km. For 20 and 50 km of fibers, a 10 Gb/s OOK channel was used to co-propagate with the QKD channel. For 100 km fiber, one, six, and eight 100 Gb/s QPSK channels were time-interleaved with the QKD channel. The synchronization channel is omitted in the experiment since a 25-MHz optical clock signal has less than -10 dBm launch power and negligible contribution to SpRS noise.

A dense wavelength-division multiplexer (Mux) with a 100-GHz grid multiplexes the quantum and classical channels at Alice's site. After co-propagation through the fiber, they are separated by a demultiplexer (DeMux) at Bob. Furthermore, the SpRS noise from classical channels is eliminated by both spectral filtering and temporal gating. A narrowband 20-GHz filter implemented by a wavelength-selective switch (WSS) is used to block the out-of-band noise. Four SPDs work in the Geiger mode with a 4-ns gate width to eliminate the out-of-window noise counts.

The experimental parameters are summarized in Table~\ref{tab:expPara}. The SPDs have a detection efficiency of $\mathrm{\eta_D=20\%}$ and a dead time of 10 $\mathrm{\mu s}$. The dark count rate is 150 counts per second or $\mathrm{Y_0=6\times10^{-6}}$ per gate. A beam splitter (BS) selects the measurement basis randomly, and then two groups of PBS take measurements in rectilinear and diagonal bases, respectively. During the key sifting, Alice and Bob compare their bases and discard those bits prepared and measured in different bases. The optical misalignment of the QKD system is $\mathrm{e_{mis}=0.5-1\%}$. To fight against the photon number splitting (PNS) attack, three pulse intensities are used~\cite{Hwang_PRL_03, Toronto_PRL_05, XB_Wang_PRL_05}, with mean photon numbers per pulse of $\mathrm{\mu_1=0.85}$, $\mathrm{\mu_2=0.04}$, and $\mathrm{\mu_3\approx0.001}$ for signal, decoy, and vacuum states, respectively. Their emission probabilities are $\mathrm{P_{\mu_1}=0.9}$, $\mathrm{P_{\mu_2}=P_{\mu_3}=0.05}$. 

\begin{table}
    \centering
    \caption{Experimental parameters}\label{tab:expPara}
    \begin{tabular}{l | l || l | l}
        \hline
        $q_X$, $q_Z$ & 0.94, 0.06 & $\eta_D$ & 0.2 \\
        $\mu_1$, $\mu_2$, $\mu_3$ & 0.85, 0.04, $\sim$0.001 & Dead time & 10 $\mu$s \\
        $P_{\mu_1}$, $P_{\mu_2}$, $P_{\mu_3}$ & 0.9, 0.05, 0.05 & $Y_0$ & $6\times10^{-6}$ \\
        $\Delta\tau$ & 4 ns & $e_{mis}$ & 0.5-1\% \\
        $\Delta\nu_B$ & 20 GHz & $f_{EC}$ & 1.2 \\
        \hline
    \end{tabular}
\end{table}

\begin{figure*}[ht]
    \centering
    \includegraphics[width=0.75\linewidth]{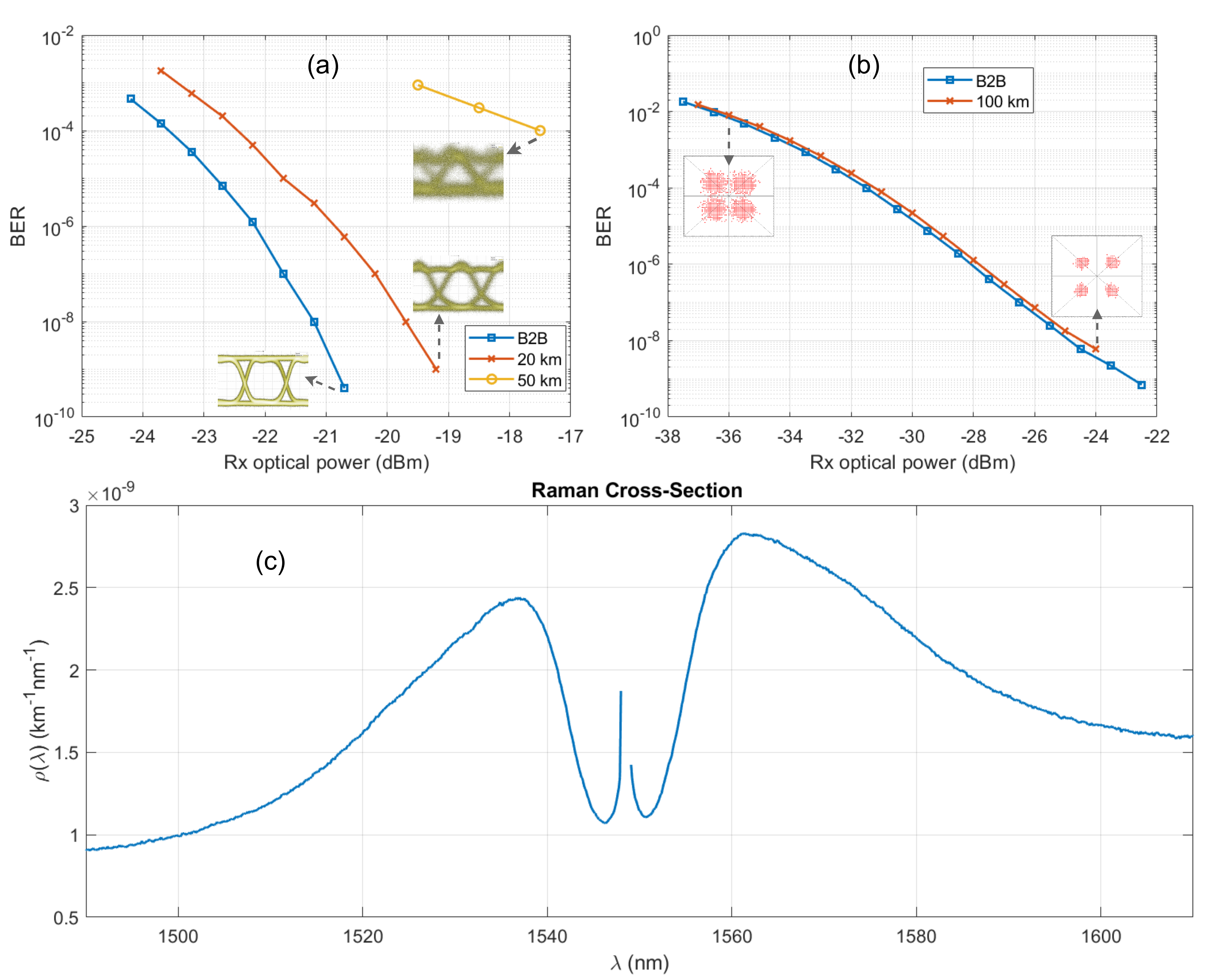}
    \caption{BER performance of classical channels and Raman cross-section. (a) BER of 10 Gb/s OOK. (b) BER of 100 Gb/s QPSK. (c) Raman cross-section of a classical channel at Ch36 (193.6 THz, 1548.52 nm).} \label{fig:classical}
\end{figure*}

Since the time-interleaving technique allows continuous QKD operation without interruption or downtime, the key size can be arbitrarily long. We followed the secure key rate estimation in~\cite{Ma_PRA_05} in the asymptotic limit of infinite key size, as shown in Eq.~\ref{eq:SKR}. R is the secure key rate in bits per pulse. q depends on the implementation of BB84 protocols and is 0.5 in our case since Alice and Bob use the same bases for half of the time. $\mathrm{ Q_{\mu_1} }$ and $\mathrm{ E_{\mu_1} }$ are the gain and QBER of signal states, and they are measured in experiments. $\mathrm{Q^L_1}$ is the lower bound of the gain of single-photon states, $\mathrm{e^U_1}$ is the upper bound of the error rate of single-photon states, and they are estimated using decoy state protocols~\cite{Ma_PRA_05}. An error correction efficiency of $\mathrm{f_{EC}=1.2}$ is used. $H_2(x) = -x\log_{2}(x)-(1-x)\log_2(1-x)$ is the Shannon binary entropy function. $\mathrm{\mu_1}$, $\mathrm{\mu_2}$, $\mathrm{\mu_3}$ and $\mathrm{P_{\mu_1}}$, $\mathrm{P_{\mu_2}}$, $\mathrm{P_{\mu_3}}$ are optimized to maximize the SKR.

\begin{equation} \label{eq:SKR}
    R \geq q\{-Q_\mu f_{EC}(E_\mu) H_2(E_\mu) + Q^L_1[ 1 - H_2(e^U_1) ] \}
\end{equation}

\section{Experimental Results}~\label{Results}

Three scenarios of quantum/classical co-propagation are tested in experiments with fiber lengths of 20, 50, and 100 km. For 20 and 50 km cases, the QKD channel was interleaved with a 10 Gb/s OOK channel. For 100 km of fiber, it co-propagates with one, six, and, eight 100 Gb/s QPSK channels. Fig.~\ref{fig:classical} shows the performance of classical channels. Due to the single-photon power level, the QKD channel causes no performance penalty to classical channels. Fig.~\ref{fig:classical}(a) shows the BER of 10 Gb/s OOK channel as a function of the received optical power with eye diagrams shown in the insets. At BER=$10^{-9}$, there is a power penalty of 1.7 dB after 20 km of fiber compared with the back-to-back (B2B) case. After 50 km of fiber, the eye diagram is severely distorted by dispersion and error-free transmission becomes impossible. Fig.~\ref{fig:classical}(b) shows the BER of a 100 Gb/s QPSK channel after 100 km fiber. Thanks to the local oscillator, the coherent receiver has a much better receiver sensitivity. Also, there is no power penalty after 100 km of fiber with the help of digital dispersion compensation. Fig.~\ref{fig:classical}(c) shows the Raman cross-section of a classical channel centered at 1548.52 nm (Ch36, 193.6 THz). To reveal the spectrum of Raman noise, the central peak at the pump wavelength is filtered out. Two local minimums are located 200-300 GHz away from the pump wavelength on both sides. The anti-Stokes noise on the shorter wavelength side is smaller than the Stokes noise on the longer wavelength side.

\begin{figure*}
    \centering
    \includegraphics[width=1\linewidth]{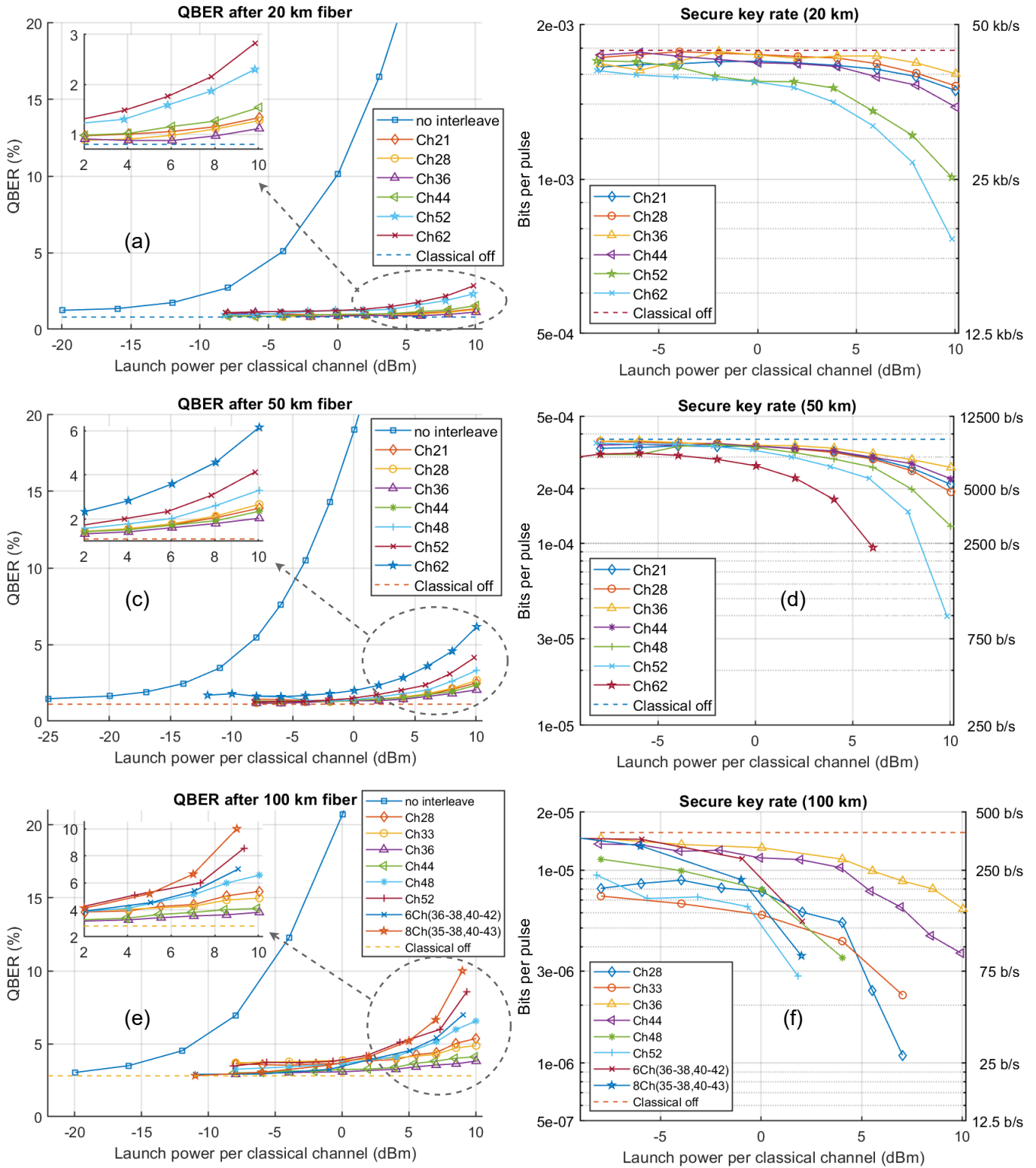}
    \caption{QBER and SKR of the QKD channel in three test scenarios. (a, b) Co-propagating with a 10 Gb/s OOK channel over 20 km of fiber. (c, d) Co-propagating with a 10 Gb/s OOK channel over 50 km of fiber. (e, f) Co-propagating with 100 Gb/s QPSK channels over 100 km of fiber.}
    \label{fig:QBER}
\end{figure*}

Fig.~\ref{fig:QBER} shows QBER and SKR as a function of the launch power of classical channels. The quantum channel is fixed at 1546.12 nm (Ch39, 193.9 THz), whereas the classical channel is tuned across the C-band to investigate the wavelength dependence of SpRS noise. The wavelength choice of the classical channel is determined by the availability of mux/demux and filters in our lab. The QBER and SKR performance of three scenarios are summarized in Table~\ref{tab:test_scenarios}. Successful QKD operation over 20, 50, and 100 km fibers with QBER less than 2.83\%, 4.14\%, and 4.33\% are demonstrated with the presence of a 10-dBm classical channel.

\begin{table*}
    \centering
    \caption{Three experimental scenarios over 20, 50, and 100 km of fibers}\label{tab:test_scenarios}
    \begin{tabular}{|cc|ccc|}
    \hline
    \multicolumn{2}{|c|}{Fiber length} & \multicolumn{1}{c|}{20 km} & \multicolumn{1}{c|}{50 km} & 100 km \\ \hline
    \multicolumn{2}{|c|}{Classical channel} & \multicolumn{2}{c|}{10 Gb/s OOK} & 100 Gb/s QPSK \\ \hline
    \multicolumn{2}{|c|}{Launch power} & \multicolumn{2}{c|}{10 dBm} & {\begin{tabular}[c]{@{}l@{}}10 dBm for 1 channel\\17 dBm for 6 channels\\18 dBm for 8 channels\end{tabular}} \\ \hline
    \multicolumn{2}{|c|}{} & \multicolumn{1}{c|}{Ch36} & \multicolumn{1}{c|}{Ch36} & Ch36 \\ \hline
    \multicolumn{1}{|c|}{\multirow{3}{*}{best}} & QBER (\%) & \multicolumn{1}{c|}{1.12} & \multicolumn{1}{c|}{2.04} & 3.81 \\ \cline{2-5} 
    \multicolumn{1}{|c|}{} & \multicolumn{1}{c|}{SKR (b/s)} & \multicolumn{1}{c|}{39500} & \multicolumn{1}{c|}{6350} & 128 \\ \cline{3-5} 
    \multicolumn{1}{|c|}{} & \multicolumn{1}{c|}{(bits/pulse)} & \multicolumn{1}{c|}{$1.58\times10^{-3}$} & \multicolumn{1}{c|}{$2.54\times10^{-4}$} & $5.1\times10^{-6}$ \\ \hline
    \multicolumn{2}{|c|}{} & \multicolumn{1}{c|}{Ch62} & \multicolumn{1}{c|}{Ch52} & Ch44 \\ \hline
    \multicolumn{1}{|c|}{\multirow{3}{*}{Worst}} & QBER (\%) & \multicolumn{1}{c|}{2.83} & \multicolumn{1}{c|}{4.14} & 4.33\\ \cline{2-5} 
    \multicolumn{1}{|c|}{} & SKR (b/s) & \multicolumn{1}{c|}{18000} & \multicolumn{1}{c|}{640} & 61.4 \\ \cline{3-5} 
    \multicolumn{1}{|c|}{} & (bits/pulse) & \multicolumn{1}{c|}{$7.19\times10^{-4}$} & \multicolumn{1}{c|}{$2.56\times10^{-5}$} & $2.45\times10^{-6}$ \\ \hline
    \end{tabular}
\end{table*}

Fig.~\ref{fig:QBER}(a) and (b) show the BER and SKR performance of a QKD channel co-propagating with a 10 Gb/s OOK channel over 20 km fiber. The SKRs are shown in terms of bits per pulse and bits per second as they are limited by the low detection efficiency (20\%) and long dead time ($\mathrm{10 \mu s}$) of our low-cost SPDs. The dashed lines show the best performance without SpRS noise when the classical channels are turned off. Without time-interleaving, QBER increases rapidly with the power of the classical channel and reaches 10\% when the classical power is 0 dBm. Beyond this power level, no QKD operation is allowed.

With the help of time-interleaving, the QBER is kept low with the presence of high-power classical channels since the SpRS noise is out of the gap window. A zoom-in is shown in the inset. The best QKD performance of QBER=1.12\% and SKR=39.5 kb/s is achieved when the classical channel is at Ch36 (193,6 THz, 1548.52 nm). This is because the QKD channel (Ch39, 193.9 THz) is at 300 GHz higher than the classical channel and within the minimum of the SpRS noise spectrum. Classical channels at Ch21 and 28 give a similar performance as Ch36; whereas classical channels at Ch52 and 62 make more SpRS noise and have worse performance. This is because anti-Stokes noise is smaller than Stokes noise. Classical channels with longer wavelengths contribute less noise than shorter wavelengths. The worst QKD performance is achieved when the classical channel is at Ch62, which is because Ch62 has the shortest wavelength and is the farthest away from the QKD channel. With a 10 dBm launch power of the classical channel, we can keep the QBER below 2.83\% and SKR above 18 kb/s for all classical channels across the C-band. Compared with other results of C-band coexistence~\cite{Univ_Geneva_NJP_10, Toshiba_PRX_12, Toshiba_APL_14, Toshiba_OE_14, Toshiba_SciRep_16}, our classical power is at least one order of magnitude higher.

Fig.~\ref{fig:QBER}(c) and (d) show the QKD performance when co-propagating with a 10 Gb/s OOK channel over 50 km fiber. The best performance of QBER 2.04\% and SKR 6.35 kb/s is achieved by placing the classical channel at Ch36. The worst performance of QBER 4.14\% and SKR 640 b/s is achieved with the classical channel at Ch52. For the classical channel at Ch62, its strong SpRS noise makes secure key generation impossible when its launch power reaches 10 dBm.

Fig.~\ref{fig:QBER}(e) and (f) show the QKD performance with 100 Gb/s QPSK channels over 100 km of fiber. Three cases are tested where the QKD channel is interleaved and co-propagates with one, six, and eight classical channels, respectively. For the single channel case, only Ch36 and 44 allow secure key generation at the classical launch power of 10 dBm. Channels at other wavelengths are too far away from the quantum channel and their SpRS noise sneaks into the gap window after 100 km of fiber. This is caused by the dispersion walk-off between classical and quantum channels. To reduce the SpRS noise and minimize dispersion walk-off, classical channels neighboring to the quantum channel (Ch39) are used, i.e. Ch36, 37, 38, 40, 41, 42 for the six-channel case, and Ch35-38 and Ch40-43 for the eight-channel case. In both cases, classical launch power is 9 dBm per channel with a total power of 16.8 dBm for the six-channel case and 18 dBm for the eight-channel case. Thanks to time-interleaving, SpRS noise from classical channels is kept out of the gate window. QBERs less than 7\% and 10\% are achieved with the presence of six or eight co-propagating classical channels.

\section{Dispersion Walk-off}~\label{Dispersion}

\begin{figure*}
    \centering
    \includegraphics[width=1\linewidth]{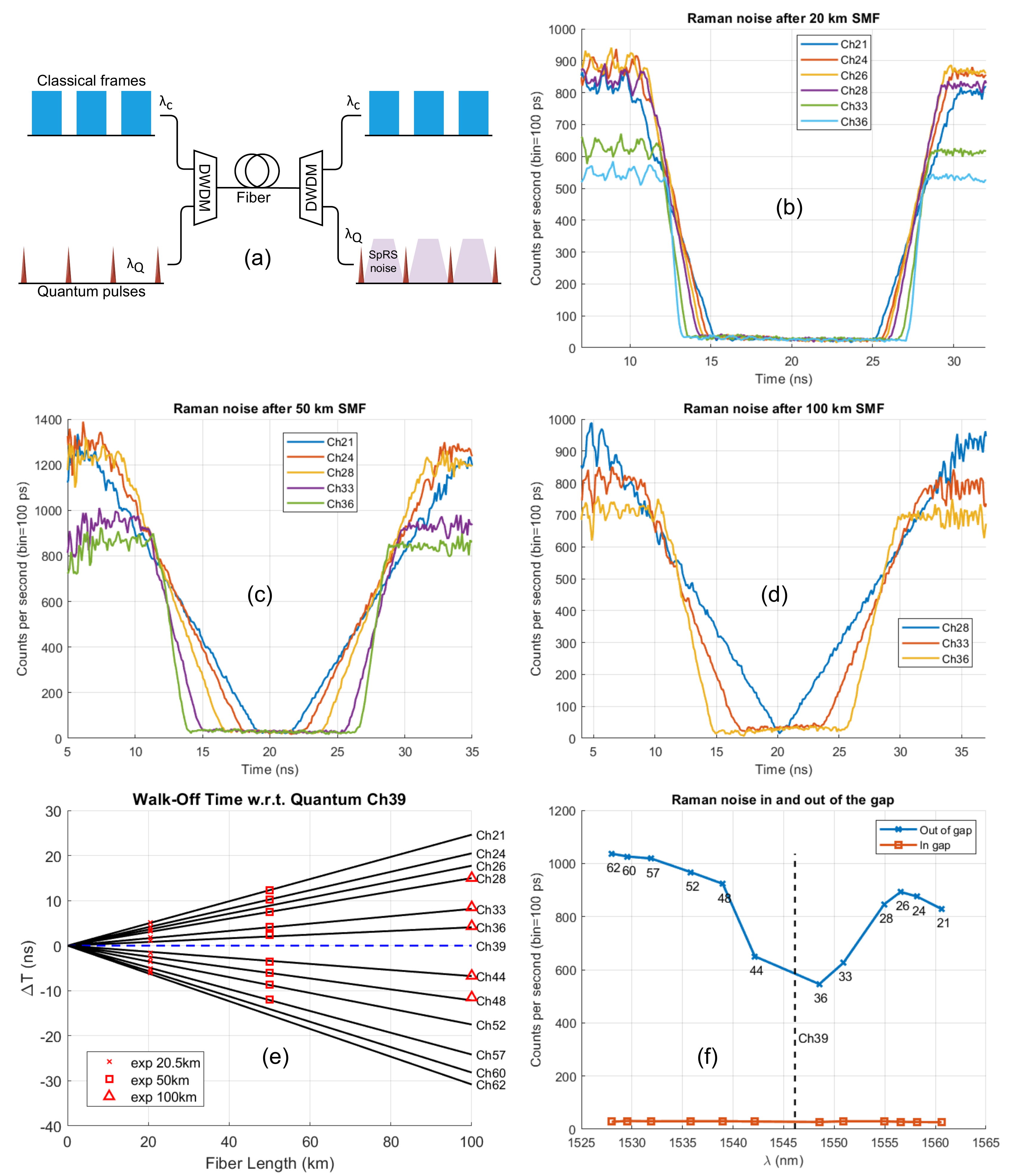}
    \caption{Dispersion walk-off between classical and quantum wavelengths. (a) SpRS noise photons generated at different locations in the fiber arrive synchronously at the fiber end. The gap windows between data frames shrink after fiber propagation of 20 km (b), 50 km (c), and 100 km (d). (e) Dispersion walk-off with respect to the QKD channel (Ch39). (f) Raman noise inside and outside of the gap window.}
    \label{fig:slope}
\end{figure*}

Fig.~\ref{fig:slope}(a) shows the dispersion walk-off between classical and quantum channels. On the transmitter side, QKD pulses are embedded into the gaps between classical data frames. A wide gap window of 15 ns is used here to allow investigation on dispersion walk-off. In real deployment, a much narrower gap window can be used. The classical and quantum channels are on different wavelengths, $\lambda_C$ and $\lambda_Q$, respectively. As the classical channel propagates along the fiber, spontaneous Raman scattering converts the incident photons from $\lambda_C$ to $\lambda_Q$. Suppose a photon is converted at the fiber distance d, it travels at the speed of $\lambda_C$ for distance d and at the speed of $\lambda_Q$ for the rest of the fiber L-d. Since different wavelengths travel at different speeds in the same fiber, noise photons generated at different locations will arrive asynchronously at the end.

Although classical data frames have steep edges, after fiber propagation, their SpRS noise is spread in the time domain. As shown in Fig.~\ref{fig:slope}(b,c,d), the noise frames after fiber propagation have gradual slopes and trapezoidal waveforms. Dispersion walk-off depends on two factors, the fiber length and wavelength difference between two channels. A longer fiber and larger wavelength difference lead to larger walk-offs. Fig.~\ref{fig:slope}(b) shows the noise gap after 20 km of fiber. Without loss of generality, we only show the classical channels with longer wavelengths than the quantum channel (Ch39), i.e., Ch21, 24, 26, 28, 33, and 36. Since Ch21 is the farthest away from Ch39, its noise frames have the least steep slopes and smallest gaps between noise frames. Ch36 is closest to Ch39, so its noise frames have the least spreading and widest gaps. 

Fig.~\ref{fig:slope}(c,d) show the noise gaps after 50 and 100 km of fibers. Longer fibers lead to more spreading of the SpRS noise, making less steep slopes and narrower gaps between noise frames. After 100 km of fiber, the noise gap is almost closed for the classical channel at Ch28 and the noise sneaks into the gate window of QKD pulses. This is why for 100 km fiber, only classical channels at Ch36 and 44 allow SKR generation, as shown in Fig.~\ref{fig:QBER}(e) and (f). In Fig.~\ref{fig:slope}(e), black lines show the calculated walk-off time of classical channels with respect to the quantum channel. The red dots are the time lengths of slopes in Fig.~\ref{fig:slope}(b,c,d) measured experimentally.

To validate the effectiveness of noise suppression in the gap windows, we measured the SpRS noise counts in and out of the gap window. Fig.~\ref{fig:slope}(f) shows the noise counts in and out of the gap window, measured with a time resolution of 100 ps. The noise counts out of the gap window show a similar wavelength dependence with the Raman cross-section in Fig.~\ref{fig:classical}(c). The noise counts inside the gap window are kept low thanks to the time-interleaving technique.

In Fig.~\ref{fig:classical}(c), SpRS noise from a classical channel is minimized at 200-300 GHz away from the pump wavelength. So classical channels should be placed close to the quantum channel to minimize their contribution to SpRS noise. Furthermore, this will also minimize the dispersion walk-off between classical and quantum channels, so that the noise photons will not sneak into the time windows for QKD pulses. In our experiments, we demonstrate 100-km co-propagation with eight classical channels surrounding the quantum channel with up to 18 dBm launch power. 

\section{Conclusions}~\label{Conclusions}

In conclusion, we present the time-interleaving technique to enable the C-band co-propagation of classical and QKD channels over 100 km of fiber. By embedding the QKD pulses into the gaps between classical data frames, we can isolate QKD pulses from SpRS noise in both wavelength and time domains. Co-propagation of a polarization-encoding decoy-state BB84 QKD channel with a 10 Gb/s OOK or 100 Gb/s QPSK classical channel in the C-band of the same fiber is experimentally demonstrated over 20, 50, and 100 km. With the presence of a 10-dBm classical channel, successful QKD operations are demonstrated over 20, 50, and 100 km of fibers with QBER of 1.12\%, 2.04\%, and 3.81\%, and secure key rates of 39.5 kb/s ($1.58\times10^{-3}$ bits/pulse), 6.35 kb/s ($2.54\times10^{-4}$ bits/pulse), and 128 b/s ($5.1\times10^{-6}$ bits/pulse), respectively. Co-propagation with six and eight classical channels with up to 16.8 dBm and 18 dBm launch power is also demonstrated with QBERs less than 7\% and 10\%, respectively.

In our experiments, the 100 km co-propagation fiber distance is limited by the low detection efficiency of the low-cost SPDs. With imrpved detection efficiency, both key rate and fiber distance can be increased. Compared with reported O-band coexistence works, our method leverages the low fiber loss in the C-band to enable longer QKD distances whereas most O-band results are bound to 80 km of fiber due to the high fiber loss. Meanwhile, compared with other C-band coexistence works, our method removes the power limit on classical channels and our launch power of classical channels is at least one order of magnitude higher than other results. Dispersion walk-off between classical and quantum channels is experimentally investigated. It is concluded that the best spectrum to place classical channels is within 200-300 GHz around the quantum channel, where both Raman noise and dispersion walk-off effect are minimized. 

The proposed C-band time-interleaving technique only handles the co-propagation scenario but cannot be used for the counter-propagation case, since the back-scattering noise photons arrive asynchronously with respect to QKD pulses. It would be a good solution to quantum/classical coexistence for long-distance ($>100$ km) point-to-point applications, e.g. metropolitan networks, where two fibers are available between every two nodes, one for each direction. For short-distance ($<80$ km) point-to-multipoint applications, such as passive optical networks (PON), due to the fiber deficiency, the QKD channel has to coexist with bi-directional classical channels, and moving the quantum channel to the O-band would be a better choice. For ultra-long distances (more than several hundred km), relay technologies or satellite QKD will be needed.

\section*{Acknowledgment}
The authors would like to thank Dr. Zhensheng Jia and Dr. Haipeng Zhang of CableLabs for the helpful discussion and for providing coherent transceivers.

\bibliographystyle{IEEEtran}
\bibliography{ref}

\end{document}